\documentstyle[12pt]{article}
\def \F{\phi}

\def\NP{{\it Nucl. Phys.\ }}

\def\PL{{\it Phys. Lett.\ }}
\def\PR{{\it Phys. Rev.\ }}

\def\b{\beta}
\def\a{\alpha}

\def\half{{1\over 2}}
\def\d{\dagger}
\def\be{\begin{equation}}
\def\eq{\end{equation}}
\def\Tr{{\rm Tr}}

\def\q2{{\rm QCD}_{2}}
\def\q4{{\rm QCD}_{4}}

\begin{document}

\begin{flushright}
OUTP-9518P\\
hep-lat/9505009\\
\end{flushright}
\vspace{20mm}
\begin{center}
{\LARGE Adjoint 2D QCD and Pure 3D QCD:\\
A Comparison of Spectra\\}
\vspace{30mm}
{\bf F. Antonuccio and S. Dalley\footnote{Based upon talk given at
the {\em Spring Workshop on String Theory, Gauge
Theory, and Quantum Gravity}, Trieste, 5-7 April 1995}}\\
\vspace{5mm}
{\em Department of Physics, Theoretical Physics\\
1 Keble Road, Oxford OX1 3NP, U.K.}\\
\end{center}
\vspace{30mm}

\abstract
We briefly review the $1+1$-dimensional $SU(N)$ gauge theory minimally
coupled to an adjoint scalar, based upon a dimensional reduction of
$2+1$-dimensional pure gauge theory, which approximates
the dynamics of the transversely polarized gluon.
The lightest glueball states are investigated non-perturbatively using
light-front quantisation in the large-$N$ limit \`{a} la 't Hooft, and
the physical meaning of the results elucidated.
Comparison is made with recent lattice Monte Carlo results for
$3$-dimensional quenched QCD.

\newpage
\baselineskip .25in
\section{Introduction}

It has been suggested by a number of authors \cite{bpr,dk,bdk1}
that adjoint scalar fields
coupled
to $1+1$-dimensional QCD provide a way of approximating in tractable
form the features of boundstates arising from
transversely polarized gluons in higher dimensional
gauge theories\footnote{Related work can be found in refs.\cite{ferm}.}.
Here we will re-examine in more detail using light-front
quantisation the low-lying glueball
states for the model describing the $k_T = 0$ gluons in pure 2+1-dimensional
gauge theory\footnote{The extension to
3+1-dimensional gauge theory with quarks will be the subject of a more
extensive forthcoming publication.} in the large-$N$ limit,
making comparison with recent preliminary
results from lattice Monte Carlo  simulation of
3-dimensional quenched QCD \cite{mike}.
Our aim is to assess the limitations of this
approach in reproducing actual features of the higher dimensional
theory. In the reduced model,
apart from an  expansion of the scale for  ratios of
boundstate masses, presumably due to  level mixing  as a result of broken
rotational invariance,
we find that the level ordering in the charge conjugation
$C=-1$ sector is a poor match with the lattice results,
while the $C=+1$ sector fares much better and is qualitatively in
agreement. We attribute the extreme precision of the
 valence gluon approximation in
the model to  transverse (physical) gluon interactions having to
proceed via an instantaneous  longitudinal intermediate gluon.

Starting from $SU(N)$ Yang-Mills theory in $2+1$-dimensions
\be
S = -{1\over 4 g_3^2} \int d^3 x \ \Tr F_{\mu\nu} F^{\mu\nu}
\eq
in the light-front gauge $A_- = 0$ one has in light-front co-ordinates
$x^{\pm} = (x^0 \pm x^1)/\sqrt{2}$
\begin{eqnarray}
S & = & {1 \over g_{3}^{2}} \int dx^2 dx^+ dx^- \Tr \left(
\partial_{+} A_2 \partial_{-} A_2  + \half (\partial_{-} A_{+})^2 + A_{+}
J^{+} \right) \label{fixed} \\
J^{+}_{ij} & = & {\rm i}[A_2, \partial_{-} A_2]_{ij} -\partial_{-}
\partial_{2}
A_{ij}^{2} \ ; \ i,j = 1,\ldots,N \label{current}
\end{eqnarray}
The field $A_{+}$ is a constrained variable which does not propogate
in light-front time $x^+$, leaving only the transverse gluon $A_2$ as
a physical degree of freedom. Much of the complexity of 3D gauge
theory is due to the linear term in the longitudinal momentum current
(\ref{current}), so one might as a first approximation  study the
theory
restricted to zero modes $\partial_{2} A_{\mu} =0$ only. This is
equivalent
to the $1+1$-dimensional adjoint gauge theory
\be
S_{R} = \int dx^0 dx^1 \Tr\left[ \half D_{\a} \F D^{\a} \F - {1\over
4g^2}
F_{\a\b}F^{\a\b} + \half m_{0}^{2} \F^2 \right] \label{trunc}
\eq
where $g^2 = g_{3}^{2} / \int dx^2$, $D_{\a} = \partial_{\a} + {\rm i}
[A_{\a}, .]$, and $\F = A_{2}/g$.
A bare mass term  has been added to (\ref{trunc}) by hand in order to
subtract a logarithmic  ultraviolet divergence later in the
calculation. $S_R$ inherits a subgroup of the Poincar\'{e}
and gauge symmetries of the $2+1$-dimensional theory. The residual
internal symmetries of $S_R$ resulting from gauge symmetries
of $S$ are just the $1+1$-dimensional gauge symmetries
\be
A_{\a} \to UA_{\a}U^{\d} + {\rm i}(\partial_{\a}U)U^{\d} \ , \ \F \to U\F
U^{\d}
\eq
$S_R$ acquires  the $SO(1,1)$ subgroup of $SO(2,1)$ Lorentz
symmetries of $S$ and also has $Z_2$ symmetries:
charge conjugation $C$ ($A_{ij}^{\mu} \to -A_{ji}^{\mu}$)
induces a discrete symmetry $\F_{ij} \to - \F_{ji}$ on the $1+1$-scalar;
there also remains the one-dimensional parity $P_1$ symmetry
($x^1 \to -x^1$).
We will in addition  make use of other dynamical information in trying to
match the $1+1$-dimensional eigentates forming represenations of the
above discrete symmetries with the representations $J$ of the full
spatial rotation group $SO(2)$ in $2+1$ dimensions.

\section{Boundstate Equations.}

When analysing the boundstates of $S_R$ by light-front quantisation,
taking the
large-$N$ limit results in additional simplifications, the calculation
being similar to, but much richer than, the one performed by 't Hooft
for quarks in the fundamental representation \cite{hoof}.
The boundstate
problem can be expressed entirely in terms of the Fourier modes
$a_{ij}(k^+)$  of the transverse gluon $\F_{ij} (x^-)$
(further
details are given in refs.\cite{dk,bdk1}). The physical Hilbert space is the
light-front Fock space constructed from the creation operators
$a^{\d}(k^{+})$. Due to confinement and the
large-$N$ limit the light-front wavefunctions one needs to consider are the
subset of singlet states of the form
\be
|\Psi (P^{+}) > = \sum_{n=2}^{\infty} \int_{0}^{P^+} dk_{1}^{+}\ldots
dk_{n}^{+} \delta \left(\sum_{m=1}^{n}k_{m}^{+} -P^{+} \right)
{f_{n} (k_{1}^{+},\ldots,k_{n}^{+}) \over N^{n/2}}
 \Tr[a^{\d}(k_{1}^{+}) \cdots
a^{\d}(k_{n}^{+})] |0> \label{wf}
\eq
Such states are eigenstates of $P^{+}$,  so the solution of the
boundstate problem is reduced to the diagonalisation of $P^{-}$ in
this basis, the dispersion relation being $M^2 \Psi= 2P^{+}P^{-} \Psi$. The
coefficients $f_n$ of $n$-gluon states, cyclically symmetric in their
arguments, satisy an infinite  set of coupled Bethe-Salpeter integral
equations \cite{bpr,dk,bdk1}
\begin{eqnarray}
{M^2 \pi \over g^2 N} f_n (x_1,x_2,\ldots , x_n) & = & {m^2\pi \over
g^2 N} {1 \over x_1} f_n (x_1, x_2, \ldots ,x_n) + {\pi \over
4\sqrt{x_1 x_2}} f_n (x_1, x_2, \ldots , x_n) \nonumber \\
&& + \int_{0}^{x_1 +x_2} dy\ \{ E[x_1,x_2,y] f_n (x_1 ,x_2,, \ldots,
x_n) \nonumber \\
&& + (D[x_1,x_2,y] - E[x_1,x_2,y]) f_n (y,x_1 +x_2 -y,x_3, \ldots, x_n) \}
\nonumber \\
&& + \int_{0}^{x_1} dy \int_{0}^{x_1-y}dz\ \{ F[x_1,y,z] f_{n+2} (y,z,x_1
-y
-z,x_2,\ldots ,x_n)\nonumber \\
&&+ F[x_3,-x_2,-x_1]
f_{n-2}(x_1+x_2+x_3,x_4,\ldots,x_n) \} \nonumber \\
&& + {\rm cyclic} \ {\rm permutations} \ {\rm of} \  (x_1,x_2,\ldots,x_n)
\label{bs}
\end{eqnarray}
\begin{eqnarray}
D[x_1,x_2,y] & = & {(x_2 - x_1)(x_1 + x_2 -2y) \over 4(x_1 + x_2)^2
\sqrt{x_1 x_2 y (x_1 + x_2 +y)}} \\
E[x_1,x_2,y] & = & {(x_1+y)(x_1 + 2x_2 -y) \over 4 (x_1 -y)^2
\sqrt{x_1 x_2 y (x_1 +x_2 -y)}} \\
F[x_1,y,z] & = & {1 \over 4\sqrt{x_1 y z (x_1 -y-z)}}\left(
{(x_1 +y)(x_1 -y-2z) \over (x_1 -y)^2} + {(2x_1 -y-z)(y-z) \over
(y+z)^2} \right)
\end{eqnarray}
where the longitudinal momentum fractions $x_m = k_{m}^{+}/P^{+}$ have
been introduced.  To these equations we must add a number of provisos.
As with the 't Hooft model \cite{hoof} there are some bare
mass ambiguities, since $\F$ has
logarithmic
and linearly divergent self-energies in $1+1$ dimensions.
The mass of an isolated gluon in higher dimensions does not get perturbatively
renormalised as a consequence of gauge invariance, but is presumably
pushed to infinity by non-perturbative confining effects. In $1+1$
dimensions on the other hand
colour flux is forced into tubes  by the restricted space
and confinement occurs perturbatively. The transverse
gluon field $\F$ has
divergent perturbative mass renormalisations.
In singlet combinations of these gluons
however, linear divergences cancel, resulting in the principal value
nature of the (Coulomb potential) $E$ integral in (\ref{bs}). For
such $1+1$-scalar particles the logarithmic divergences do not cancel and one
must decide what to do with them. Adding a bare mass $m_{0}$ to
$S_R$, which does not violate the reduced gauge invariance,
we will adopt the
prescription of ref.\cite{bdk1} setting the renormalised
 mass $m=0$ in (\ref{bs}).
This is clearly not the only prescription one could take, but is perhaps
intuitively closest to the three dimensional theory,
which one is trying to model as accurately as
$1+1$-dimensions allow, since the spectrum is unbounded for $m<0$
while for $m \to 0^+$ a positive spectrum is given in units of the only
dimensionful scale $g^2$. More generally one might keep it as a free
parameter
and fit to a known spectrum in the spirit of ref.\cite{bpr}.
Our choice of a perturbative (light-front) vacuum $|0>$ in any case
implies that we are neglecting  zero modes in this
two-dimensional
model --- we see no reason or evidence for zero modes  condensing
in this case  --- the perturbative vacuum being sufficient for
confinement in $1+1$ dimensions.

Simple approximate analytic solutions may be found to the
equations (\ref{bs})  while numerical
solutions may be easily obtained to any desired accuracy with the help of
Mathematica and a workstation. By cutting the interval of allowed momentum
fractions $0<x<1$ into $x \in (1/K,3/K,5/K,\cdots)$ for some integer
$K$, the problem becomes  one of finite matrix diagonalisation
\cite{bhp}, the continuum limit being achieved by extrapolating $K\to
\infty$.
We found it particularly useful to employ a Lanczos algorithm in
computing and diagonalising the mass matrix $M^2$ in this way.
A Hilbert space of dimension $\sim O(10^3)$ can be comfortably
handled.
The elementary processes (fig. 1) for adjoint particles represented
in (\ref{bs}) include
the linear Coulomb potential ($E$ integral),
present also for fundamental representation
particles but with doubled strength $g^2 \to 2 g^2$ since an adjoint
source has two flux lines attached to it rather than one. The
glueballs
(\ref{wf}) may therefore be
pictured as a ring of flux created by a closed chain
of
gluons. In addition in (\ref{bs}) there is a $2 \to 2$
annihilation channel ($D$ integral) and pair
creation and annihilation  of gluons ($F$ integral), which are
not suppressed by the large-$N$ limit but for low-lying levels are
kinematically suppressed \cite{dk}.
A survey of the broad features of the mass spectrum $M^2$ resulting
from adjoint scalars in 2D has been performed
in refs.\cite{bpr,bdk1}. A massive spectrum of stable glueballs is organised
into approximate  valence gluon trajectories for low-lying levels, while at
higher energies pair creation dominates to give a complicated picture.
We have concentrated on the highly structured eight or so lowest states
resulting
from (\ref{bs}) in
order to compare with  Monte Carlo results of the unreduced
3-dimensional lattice
theory. In addition to measuring the masses $M$ from extrapolation of
finite $K$ calculations to the continuum limit $K =\infty$, we also
used the light-front wavefunctions to compute
structure functions of these mass eigenstates at fixed
cut-off.
In particular the quantities
\begin{eqnarray}
<n> & = & \sum_{n=2}^{\infty} n
\int_{0}^{1} dx_{1}\ldots dx_{n} \delta \left(\sum_{i=1}^{n}
x_{i} -1 \right) |f_{n}(x_1 , \ldots ,x_n )|^2 \ , \\
g(x) & = & \sum_{n=2}^{\infty} \int_{0}^{1} dx_{1}\ldots dx_{n} \delta
\left(\sum x_i -1\right) \sum_{i=1}^{n} \delta ( x_i - x) |f_n|^2
\end{eqnarray}
give the average number  of  gluons and the number of gluons
with momentum fraction between $x$ and $x+dx$ in a boundstate. These
quantities are helpful in classifying states according to expected
quantum numbers of the unreduced $2+1$-dimensional theory.

\section{Solutions.}

The  numerical solutions for the  reduced theory are given in table 1
and illustrated  on fig.2 and fig.3.
As in
refs.\cite{dk,bdk1}
we note that a spectacularly accurate valence gluon approximation is at work in
the light states.
The Coulomb potential, which includes the 2nd term in (\ref{bs}) as
well
as the $E$-integral, dominates since it is the
only positive definite and singular amplitude. Both the annihilation
channel ($D$) and pair production ($F$)
amplitudes take either sign with roughly
equal probability, leading to much destructive interference. Physically they
are suppressed because the intermediate $A_+$ `particle' is
non-propagating in light-cone time $x^+$, so couples unfavourably to physical
transverse gluons pairs in the boundstate.

In labeling the states with their $(P_1,C)$ quantum numbers,
a technical problem arises in determining $P_1$, which has the effect
$x^+ \leftrightarrow x^-$ on vectors. This symmetry is broken when the
momentum fractions $x$ are discretised, since this is equivalent to
(anti)periodic
conditions on $x^-$. Under $P_1$
\be
x_{m} \to {1 \over x_{m} \sum_{m'=1}^{n} {1\over x_{m'}}}
\eq
which in general is not of the form ${\rm integer}/K$. There are
various quantitative and qualitative
ways of assessing by inspection the $P_1$ of mass
eigenstate at fixed
cut-off
$K$. The qualitative one
we found particularly useful was to note that in general
$P_1$ transforms  Fock states where one gluon carries most of the
momentum to Fock states where the momentum is shared evenly between
all the gluons.

A striking property of fig.2 is the almost
linear $M$ versus $<n>$ trajectory \cite{dk,bdk1}, which is
approximately
repeated before becoming diluted by pair production effects at higher
mass. Such a trajectory  represents  `radial excitation' of the glueball
flux loop in the sense that the mass of a state is increased
by adding  gluons and their attendant flux lines to the ring.
The leading radial trajectory shows featureless structure functions
(fig.3) peaked
at $1/<n>$, while the higher trajectory exhibits further oscillatory
behaviour analogous to  higher `angular momentum' states (therefore the
Regge trajectories would run vertically in fig.2). For the latter
there is an increased probability to find an asymmetrical sharing of the
momentum between gluons in the glueball.
It is interesting to
note that $g(x)$ remains large for $x\sim 0$ in general. In particular
the approximately two-gluon states  are rather like cosine
wavefunctions, which was also found for the
low-lying
mesons in the 't Hooft model at small quark mass \cite{bhp}.
In the latter case the massless groundstate meson formed from
massless
 quark and anti-quark
had a constant wavefunction $f_2(x,1-x) = {\rm const.}$,
corresponding to
non-interacting quarks. In the present case
the approximate $(++)$ glueball groundstate wavefunction
$f_2(x,1-x) \equiv g(x)$
is not quite constant due to the
extra spin-related contribution  $\pi / 4 \sqrt{x_1 x_2}$ to its mass
in (\ref{bs}), suppressing it at the
endpoints; this gives the glueball spectrum
its mass gap.  Indeed a pretty good estimate of the groundstate mass
is afforded by first-order Rayleigh-Schrodinger
perturbation theory with $f_2(x,1-x)=1$ as zeroth-order wavefunction,
\begin{eqnarray}
M^2 \approx <g^2 N / 2\sqrt{x(1-x)} >_{f_2 =1} & = & \int_{0}^{1}
dx {g^2 N \over 2
\sqrt{x
(1-x)}} \nonumber \\
&= & {g^2 N \pi \over 2}
\end{eqnarray}
yielding $M \sim 2.2 \sqrt{Ng^2/\pi}$, differing from the numerical
solution by $5\%$. In fact all
the wavefunctions on the leading radial trajectory are well described
by free gluons $f_n(x_1,x_2,\ldots, x_n) = 1$, yielding $g(x) = 2n(1-x)$, plus
suppression at the endpoints due to the $1+1$-dimensional remnant of
spin interactions, which tend to favour a symmetrical sharing of the momentum.

\section{Comparison with Lattice Results.}

The most reliable lattice Monte Carlo data for $SU(3)$ glueball masses in three
dimensions
\cite{mike} is shown  on
fig.4. Less reliable data exists for states nominally of
 higher mass than those shown. The classification is $J^{PC}$, where
parity $P$
in two space dimensions is taken to be reflection about {\em one} axis (e.g.
$x^1 \to -x^1$, $x^2 \to x^2$).
Particles of non-zero angular momentum $J$ should be degenerate in $P$-doublets
$|J> \pm |-J>$.
Although we have no way of telling  the true strength of breaking of $SO(2)$
rotational  symmetry by the dimensionally reduced theory, which
reduces it to a $Z_2$ subgroup, we will attempt to match our states
with $J^{PC}$ labels. Firstly we take $P\equiv P_1$.
Under $180^{0}$ rotations $x^1 \to -x^1$ and  $x^2 \to -x^2$,
and although $\F$ is a $1+1$-dimensional scalar, it is the
component
of a $2+1$-dimensional vector, so we will assume this induces
$P_1$ and $\F \to -\F$. Since all the states we study are found to be
invariant
under $\F_{ij} \to \F_{ji}$, which reverses the ordering of gluons
around
a flux ring, this implies that  $CP_1 \equiv |J|\ {\rm mod2}$, which
rules out half the possible states in three dimensions from the very
beginning. In fact it removes precisely one of the states from each
$P$-doublet.
One is still left with the problem of spin labeling
within the set of even and odd spins separately, because in general
there will be mixing within each set, i.e. spin 0 mixes with spin 2
etc..
We took the spin label  for states of given $(P_1,C)$ on a radial
trajectory to be the lowest which has not appeared on a lower energy trajectory
with the same $(P_1,C)$.

The mass ratio scale gets expanded in the truncated theory, presumably
due to the mixing between levels as a result of $SO(2)$-breaking,
 so we have shrunk it
by an appropriate factor in fig.4 in order to better compare the level
ordering with lattice results.
It is tempting, in this gauge and in this large-$N$ light-front
quantisation scheme, to describe the data in terms of a
constituent gluon picture.
However, in at least one case this is in clear conflict
with the ($N=3$) lattice data: a $O^{--}$ state
could be interpreted as an orbital
excitation of the 3-gluon groundstate $1^{+-}$; on the lattice the
$O^{--}$ is the lower state since it couples to the combination
$U_{P} - U_{P}^{\d}$ of one Wilson plaquette $U_{P}$, while $1^{+-}$
needs a longer lattice Wilson loop. In fact fig.4 shows that the  ($N=\infty$)
dimensionally reduced spectra in the $C=-1$ sector predict
light spin 1 glueballs, in disagreement with the 3D lattice.
The ordering in the $C=+1$ sector is quite good by comparison.
Spin 1's being lower than spin 2's  may be due to the mixing
of even spins and
odd spins  amongst themselves since spin  1 is pushed down by
mixing with spin 3 etc.

Lattice results also exist for $SU(2)$
when $C=+1$ \cite{mike},
but states with $C=-1$  do not appear  since Wilson loops
are  unoriented.
For completeness we have therefore plotted an $N=\infty$ extrapolation
for $C=+1$ from $N=2,3$ data by fitting to the form $M_{\infty} =
M_N -{\rm const.}/N^2$, though the variation is actually quite small.
We do not think
that finite $N$ corrections are a major source of error, relatively
speaking, but that the discrepancies we have found are probably
largely  of a kinematic nature due to the severity of
the reduction and hence the smallness of
the residual $Z_2$ symmetry.
We are however encouraged by the simple physical picture that the
reduced theory presents since it  incorporates the two principal
dynamical effects at work in hadrons, the linear string potential and
spin-dependant interaction. Although the reduced model also includes
pair production of transverse gluons, the explicit computation
shows this to be suppressed
to a remarkably high degree.
There is no direct self-coupling
of transverse gluons in three dimensions, only coupling  through
a longitudinally polarized non-propagating gluon, which
we believe  is the origin of
the suppression of pair production.
The situation when reducing gauge theory from four to two dimensions is rather
different and it is clearly of interest to perform a similar analysis
for the $1+1$-dimensional two-adjoint scalar theory obtained by
ignoring $x_{T}$-dependence. It has the light-front hamiltonian
$(x^{\pm} = (x^0 \pm x^3)/\sqrt{2})$
\begin{eqnarray}
P^- & = & g^2 \int dx^- \Tr \left( - \half J^{+} {1 \over
\partial_{-}^{2}}J^{+} -  {1 \over 2} [\F_1 , \F_2]^2 \right) \\
J^{+} & = & {\rm i} [\F_1 , \partial_{-} \F_1 ] + {\rm i} [\F_{2}, \partial_{-}
\F_2 ] \ ,
\end{eqnarray}
where $\F_{i} \sim A_{i}$ are the transverse zero modes of the gauge
potential, and includes a 4-point contact interaction between
transverse
fields. The discrete remnant of the rotation group in this case
is the much larger Dihedral
group $D_4$, so there is some hope that the
mixing between different spins is  lessened.
A detailed analysis of the hadron spectrum,
including quarks, will be given elsewhere.

\vspace{10mm}

\noindent Acknowledgements: We would like to thank I. Klebanov, I. Kogan,
J. Paton, and M. Teper for very valuable discussions, and M. Teper for
making available to us his lattice results.

\vspace{5mm}
\begin{center}
FIGURE CAPTIONS
\end{center}
\noindent
Figure 1 -- The four elementary processes contributing at order $g^2$
in
$S_R$.

\noindent
Figure 2 -- Mass spectrum extrapolated to the continuum limit $K = \infty$,
classified
by $(P_1, C)$. The estimated error in these masses is a few percent,
though great accurracy is not terribly important for the purposes of
this
paper. Note that the $P_1$ eigenvalue of the highest state shown could not be
identified
with certainty (see text).

\noindent
Figure 3 -- Structure functions of mass eigenstates:
(a) $<n> \sim 2$ ; (b) $<n> \sim 3$ ;
(c) $<n> \sim 4$ ; (d) $<n> \sim 5$.
The solid lines refer to states on the lower radial trajectory in
fig.2
while the chain lines refer to states on the upper trajectory.

\noindent
Figure 4 -- Comparison of dimensionally reduced and 3D lattice
glueball mass
ratios to the groundstate.
$SU(3)$ lattice data has solid circles, the $N \to \infty$
extrapolation of lattice data where available has open circles,
and the reduced theory has squares. Note the different scales;
that in parentheses refers to the reduced
theory results.

\medskip

\medskip

\noindent
Table1 - M is the $K=\infty$ extrapolated mass using the full basis of
states.  $<n>$ is calculated from the theory truncated to
the sector of 2, 4, and 6 gluons only for $K=20$, or
to the sector of 3 and 5 gluons for $K=19$.

\medskip

\begin{tabular}{|c|c|c|c|c|c|c|c|c|} \hline
   & $0^{++}$ & $1^{+-}$ & $0_{\ast}^{++}$ & $1_{\ast}^{+-}$
       & $2^{++}$ & $0^{--}$ & $0_{\ast \ast}^{++}$ &
          $2^{++}_{\ast}/1^{-+}?$ \\ \hline
     M & 2.1 & 3.4 & 4.7 & 5.9 & 6.3 & 6.6 & 7.1 & 7.4 \\ \hline
     $<n>$ & 2.002 & 3.002 & 4.000 & 4.990 & 2.175 & 3.085 &
       5.902 & 4.039 \\ \hline
\end{tabular}

\vfil
\end{document}